\documentclass{article}

\usepackage{amsmath}
\usepackage{amsfonts}
\usepackage{amssymb}
\usepackage{amsthm}
\usepackage{complexity}
\usepackage{microtype}
\usepackage{tikz}
\usepackage{comment}
\usepackage{fullpage}
\newtheorem{theorem}{Theorem}[section]
\newtheorem{lemma}[theorem]{Lemma}
\newtheorem{corollary}[theorem]{Corollary}

\newtheorem{definition}[theorem]{Definition}

\newcommand{\xin}{x_{\textrm{in}}}
\newcommand{\yin}{y_{\textrm{in}}}
\newcommand{\zin}{z_{\textrm{in}}}
\newcommand{\xout}{x_{\textrm{out}}}
\newcommand{\yout}{y_{\textrm{out}}}
\newcommand{\zout}{z_{\textrm{out}}}

\newcommand{\ytok}[1][black]{\tikz[baseline=-0.6ex]\draw[fill=gray,radius=3pt] (0,0) circle ;}%
\newcommand{\ntok}[1][black]{\tikz[baseline=-0.6ex]\draw[fill=white,radius=3pt] (0,0) circle ;}%

\begin{document}

\title{Reconfiguration of Satisfying Assignments and Subset Sums:\\Easy to Find, Hard to Connect}
\date{}

\author{Jean Cardinal\footnote{Universit\'e libre de Bruxelles (ULB), \texttt{\protect{jcardin@ulb.ac.be}}}
\and 
Erik D. Demaine\footnote{Massachusetts Institute of Technology, \texttt{\protect{edemaine@mit.edu}}}
\and 
David Eppstein\footnote{University of California, Irvine, \texttt{\protect{eppstein@ics.uci.edu}}}
\and
Robert A. Hearn\footnote{\texttt{\protect{bob@hearn.to}}}
\and
Andrew Winslow\footnote{University of Texas Rio Grande Valley, \texttt{\protect{andrew.winslow@utrgv.edu}}}}

\maketitle

\begin{abstract}
  We consider the computational complexity of {\em reconfiguration} problems, in which one is given two combinatorial
  configurations satisfying some constraints, and is asked to transform one into the other using elementary transformations,
  while satisfying the constraints at all times. Such problems appear naturally in many contexts, such as model checking,
  motion planning, enumeration and sampling, and recreational mathematics. We provide hardness results for problems in this family, in which the constraints and operations are particularly simple.
  
  More precisely, we prove the \PSPACE{}-completeness of the following decision problems:
\begin{itemize}
\item Given two satisfying assignments to a planar monotone instance of Not-All-Equal 3-SAT, can one assignment be transformed into the other by single variable ``flips'' (assignment changes), preserving satisfiability at every step?
\item Given two subsets of a set $S$ of integers with the same sum, can one subset be transformed into the other by adding or removing at most three elements of $S$ at a time, such that the intermediate subsets also have the same sum?
\item Given two points in $\{0,1\}^n$ contained in a polytope $P$ specified by a constant number of linear inequalities, is there a path in the $n$-hypercube connecting the two points and contained in $P$? 
\end{itemize}
These problems can be interpreted as reconfiguration analogues of standard problems in \NP{}. Interestingly, the instances of the \NP{} problems that appear as input to the reconfiguration problems in our reductions can be shown to lie in \P{}. In particular, the elements of $S$ and the coefficients of the inequalities defining $P$ can be restricted to have logarithmic bit-length.
\end{abstract}

\section{Introduction}

Many computational problems consist of deciding the existence of a combinatorial object subject to constraints expressible in algebraic or logical terms. We consider {\em reconfiguration} problems, in which one is given two objects satisfying a set of constraints, and the goal is to transform one into the other using simple reconfiguration moves such that all the constraints remain satisfied at every intermediate step. Such problems find applications in dynamic environments or reactive systems, in which solutions are required or designed to evolve, in accessibility problems in model checking, as well as in enumeration and sampling problems, in which connectivity of the search space plays a major role.

We focus on reconfiguration problems that are naturally derived from standard \NP{}-complete problems. This line of inquiry seems to have begun with the Sliding Tokens problem, a reconfiguration version of Independent Set, by Hearn and Demaine~\cite{Hearn-2005a}, and has gained momentum with publications such as the extension of Schaefer's dichotomy to the connectivity of Boolean satisfiability due to Gopalan et al.~\cite{GKMP09}, and an overview of the complexity of reconfiguration problems by Ito et al.~\cite{Ito-2011a}. In the canonical example of Boolean satisfiability, one is given two satisfying assignments to connect by a sequence of variable assignment flips, such that the formula remains satisfied at every step. The study of this type of question also benefits from the interest of puzzle designers and recreational mathematicians; token-sliding problems, for instance, are related to the famous 15-puzzle, popular in the late 19th century~\cite{Slocum-2006a}. Combinatorial reconfiguration now constitutes a quickly developing field with dedicated research groups and workshops (such as the combinatorial reconfiguration workshop held in Banff in January 2017).
For a more thorough survey and history of this family of problems, we refer to van den Heuvel~\cite{vdH13}. 

Reconfiguration of independent sets in graphs is among the most studied problem in this vein (see~\cite{BKW14,B16,DDFHIOOUY15,HU16} for recent results) and is relevant to our findings. In these problems, one is given a graph $G$ and two independent sets of $G$ of the same size $k$, and the goal is to transform one into the other using elementary operations, preserving independence at every step. The operations consist either of ``token slides'', in which a vertex in the independent set is replaced by one of its neighbors~\cite{Hearn-2005a}, or of vertex additions and removals such that the size of the independent set is either $k$ or $k-1$~\cite{Ito-2011a}. A third, related, model is that of ``token jumping'', in which a vertex is replaced by another, so that the size remains unchanged~\cite{KMM12}. In general, these reconfiguration problems are known to be \PSPACE{}-complete.

Reconfiguration problems for graph colorings followed, and have a large dedicated body of results as well~\cite{BLPP14,BC09,BMNR14,CHJ11,IKD12}. Again, many such problems are known to be \PSPACE{}-complete. Reconfiguration problems for shortest paths~\cite{B13,KMM11}, vertex covers~\cite{INZ16}, dominating sets~\cite{HIMNOST16}, and Steiner trees~\cite{MIZ16} have also been considered.

% J: note about the related problems of *shortest* reconfiguration path and all-pairs connectivity?

As discussed by van den Heuvel~\cite{vdH13}, the question of the relation between the complexity of the existence problem (of a satisfying assignment, for instance) and that of the reconfiguration problem is intriguing. In many early examples, reconfiguration problems in \P{} are obtained from existence problems that are in \P{}, and many \PSPACE{}-hardness proofs follows the lines of the \NP{}-hardness proof of the corresponding satisfiability problem. In the Schaefer-type dichotomy theorem established by Gopalan et al.~\cite{GKMP09}, all satisfiability problems in \P{} yield a reconfiguration problem in \P{} as well. In some cases, the satisfiability problem is \NP{}-complete while the reconfiguration problem is in \P{}. (For example, this is the case for 1-in-3 SAT, whose reconfiguration problem is trivial.) Examples in which the existence problem is in \P{}, but the reconfiguration problem is \PSPACE{}-complete can also be found. Prominent examples are reconfiguration of shortest paths~\cite{B13} and reconfiguration of 4-colorings of bipartite and planar graphs~\cite{BC09}. 
Our results provide further examples of such a situation.

\paragraph{Our results.}
We give hardness results for reconfiguration problems involving solutions of special families of Boolean satisfiability problems, subset sum and knapsack problems, and, more generally, 0-1 linear programming problems.

In Section~\ref{sec:sat}, we prove that the problem of reconfiguring satisfying assignments to a planar monotone instance
of Not-All-Equal 3-SAT by single variable flips is $\PSPACE$-complete. Interestingly, the planar Not-All-Equal 3-SAT problem is in \P{}.
If we further restrict to monotone instances, the reconfiguration problem is equivalent to reconfiguration of 2-colorings of 3-uniform hypergraphs with planar vertex-edge incidence graphs.

In Section~\ref{sec:subset}, we consider the Subset Sum reconfiguration problem, that is, reconfiguration of subsets of a set of integers with the same sum. For this, we need to be able to perform elementary moves involving three elements of the set. We show that this problem is again \PSPACE-complete.

Finally, in Section~\ref{sec:paths}, we prove the \PSPACE-completeness of the problem of finding a path between two points of the hypercube that is constrained to lie within a polytope. We show that the hardness result holds even if the number of inequalities defining the polytope is $O(1)$, and the coefficients involved are polynomial.

\section{Planar NAE 3-SAT Reconfiguration}
\label{sec:sat}

In this section, we give new results on the reconfiguration problems for a variant of Boolean satisfiability.
%The corresponding reconfiguration problems can be cast as follows:

\begin{definition}[Boolean Satisfiability Reconfiguration Problem]
Given an instance of a Boolean satisfiability problem and two satisfying assignments $s$ and $t$, does there exist a sequence of satisfying assignments $s_1,s_2,\ldots ,s_k$ such that $s_1=s$, $s_k=t$, and for all $i \in [k-1]$, $s_{i+1}$ can be obtained from $s_i$ by a single variable flip?
\end{definition}

Such problems (also referred to as the \emph{$s-t$-connectivity} problems for Boolean satisfiability)
have been considered extensively before~\cite{GKMP09,MTY10,MTY11,S14,MNPR15,S15}. 
Here we investigate the complexity of the reconfiguration versions of Boolean satisfiability problems in which the \emph{variable-clause incidence graph} is planar. 
The variable-clause incidence graph of a CNF formula is a bipartite graph in whose set of vertices
is the union of the set of clauses and the set of variables of the formula, and a variable vertex is adjacent
to a clause vertex if the variable appears in the clause, in either positive or negative form.
The \emph{planar 3-SAT problem} is the 3-SAT problem restricted to instances with a planar variable-clause incidence graph. 
It has long been known that planar 3-SAT is \NP-complete~\cite{Lichtenstein-1982a}.

In the \emph{NAE 3-SAT problem}, satifying assignments are forbidden from containing clauses in which all literals have the same value. 
Hence in a satisfying assignment, every clause has exactly two literals with the same value. 
In an instance of \emph{Monotone NAE 3-SAT}, all literals appearing in the clauses are positive.

Monotone NAE 3-SAT is equivalent to 2-coloring 3-uniform hypergraphs, and known to be \NP{}-complete from Schaefer's dichotomy theorem.
We consider instances of \emph{Planar NAE 3-SAT}, where the variable-clause incidence graph is planar.
In 1988, Moret proved the surprising result that Planar NAE 3-SAT is in \P{} by reducing the problem to that of finding a maximum cut in a planar graph~\cite{Moret-1988a}. 
We prove:

\begin{theorem}
\label{thm:planarNAESAT}
Planar Monotone NAE 3-SAT Reconfiguration is \PSPACE-complete.
\end{theorem}

It is interesting to observe that the problem is \PSPACE{}-complete despite the satisfiability problem lying in \P{}. 
The proof relies on the Nondeterministic Constraint Logic framework of Hearn and Demaine~\cite{Hearn-2005a,HD}.

\paragraph{Nondeterministic Constraint Logic (NCL).}
In nondeterministic constraint logic, a \emph{constraint graph} is an edge- and node-weighted graph. 
A \emph{configuration} of such a graph is an orientation of its edges, and an orientation is \emph{legal} provided that the sum of the weights of edges pointing to a node is at least the weight of this node.
In what follows, we will further restrict to graphs in which all node weights equal 2, and edges have weights either 1 or 2. The latter are referred to as red and blue edges, respectively. Furthermore, we only have two types of nodes: AND nodes with one blue and two red incident edges, and OR nodes with three blue incident edges. It was proved that the framework retains all of its expressive power, even under these restrictions~\cite{Hearn-2005a}. The names of the two node types come from the interpretation of the incoming weight constraint: a configuration is legal if and only if (i) for all AND nodes, the blue edge is not outgoing unless both red edges are incoming, (ii) for all OR nodes, at least one edge is incoming.

\begin{definition}[C2C Problem]
Given a constraint graph and two legal configurations $C_1$ and $C_2$, can $C_2$ be obtained from $C_1$ by flipping one edge at a time, so that all intermediate configurations are also legal?
\end{definition}

\begin{theorem}[\cite{Hearn-2005a}]
\label{thm:c2c}
The C2C problem is \PSPACE{}-complete, even if the constraint graph is restricted to be planar.
\end{theorem}

As a warmup, we first consider the known reduction from the planar C2C problem to planar 3-SAT reconfiguration. 
Given a planar constraint graph, we define one Boolean variable per edge. 
When considering an edge $x$ incident to a node, we denote by $\xin$ the literal corresponding to the orientation of $x$ towards the node, and the opposite literal by $\xout$. 
For a given AND node with a blue incident edge $x$ and two red incident edges $y$ and $z$, we add the two clauses $(\xin \vee \yin)$ and $(\xin \vee \zin)$, forcing both $\yin$ and $\zin$ to be true whenever $\xin$ is false. 
For a given OR node with three incident blue edges $x$, $y$, and $z$, we add the single clause $(\xin \vee \yin \vee \zin)$. 
The resulting variable-clause incidence graph is planar whenever the initial constraint graph is. 
This reduction is due to Sarah Eisenstat,\footnote{MIT Course 6.890, ``Algorithmic Lower Bounds: Fun with Hardness Proofs'' (Fall '14), Lecture 17.} and is also alluded to by Gopalan et al.~\cite{GKMP09}. 

\begin{figure}
\begin{center}
\includegraphics[page=1,width=.6\columnwidth]{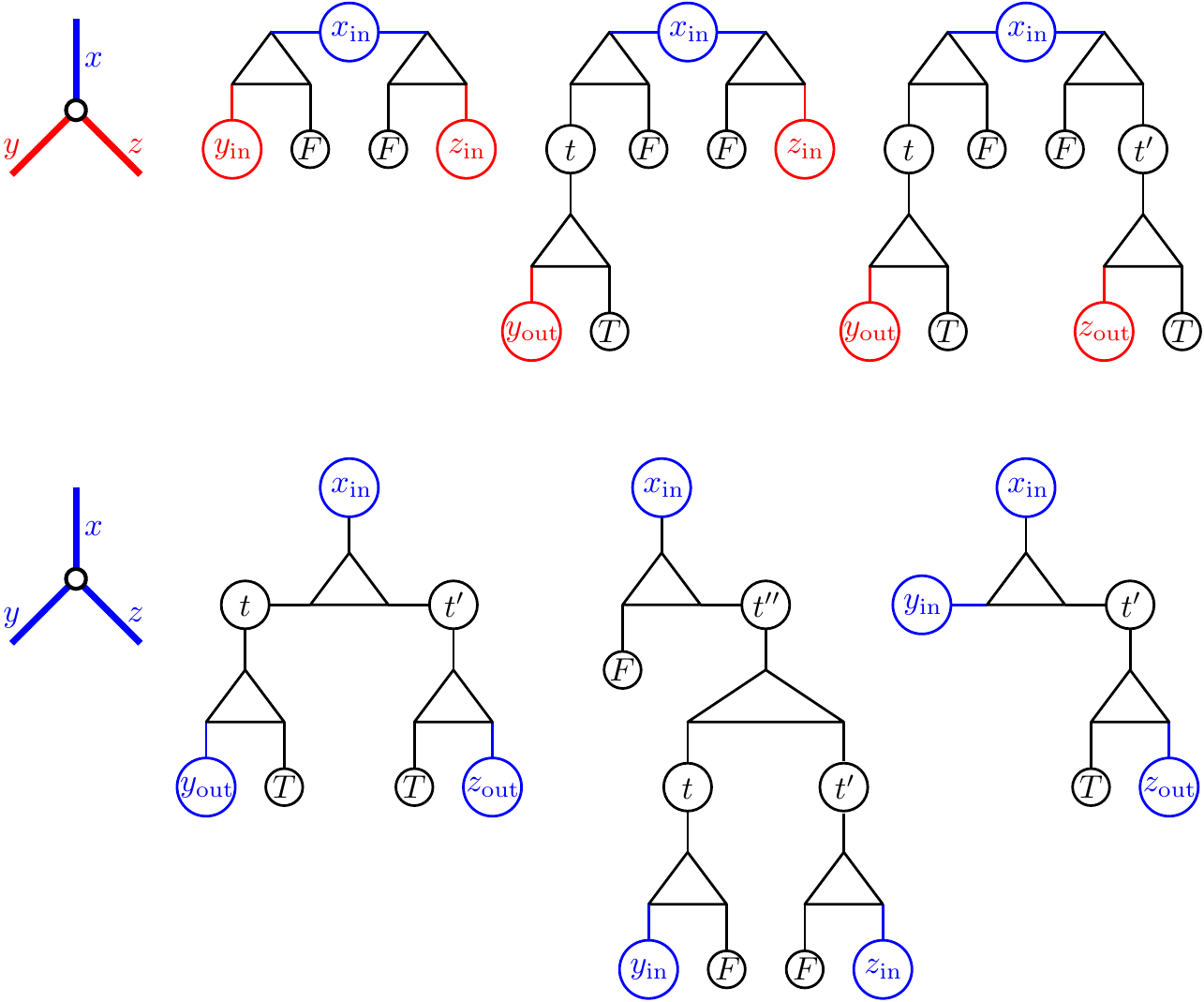}
\end{center}
\caption{\label{fig:ANDOR}Monotone NAE clauses implementing the AND (top) and OR (bottom) nodes in a constraint graph. 
%The AND gadgets force both $\yin$ and $\zin$ to be true whenever $\xin$ is false.
%OR gadgets forbid exactly one triple of orientations, in which all three edges are outgoing.
Only one of the three versions of the gadget is used, depending on the encoding used for the edge orientations.}
\end{figure}

\begin{comment}
The hardness proof of Theorem~\ref{thm:planarNAESAT} is more involved. 
The AND and OR nodes are also translated into monotone NAE clauses involving variables representing edges of the constraint graph, but also other additional variables. 
Furthermore, the translation may use different sets of clauses depending on whether the variable representing the edge corresponds to the incoming or outgoing orientation. 
The reduction is summarized on Figure~\ref{fig:ANDOR}. 
In this figure, triangles represent monotone NAE clauses. 
The symbols $F$ and $T$ represent variables whose values have been fixed by a ``rigid'' gadget. 
\end{comment}

\begin{proof}[Proof of Theorem~\ref{thm:planarNAESAT}]
Membership in \PSPACE{} can be proved by exhibiting a nondeterministic polynomial space algorithm and applying Savitch's Theorem.
To prove \PSPACE-hardness, we reduce from the planar C2C problem by implementing the two types of nodes with Monotone NAE clauses.

We first describe a set of NAE clauses together with a satisfying variable assignment such that no variable can be flipped.
This will allow us to set a variable to a certain Boolean value.
We use the following four monotone NAE clauses:
\begin{equation}
  \label{gadget:rigid}
(t \vee x \vee z) \wedge (t \vee y \vee z) \wedge (t \vee x \vee y) \wedge (x \vee y \vee z). 
\end{equation}
Now we set $t\gets \mathrm{true}, z\gets \mathrm{true}, x\gets \mathrm{false}, y\gets \mathrm{false}$. 
The clauses and the assignment are illustrated in Figure~\ref{fig:rigid}. In this figure and the following, the triangles represent the clauses.
It can be checked that every variable is contained in a clause in which no other variable is set to the same value. Therefore, this assignment is isolated in the reconfiguration graph, and can be used to set the value of a variable.  
Recall that an instance of the Boolean satisfiability reconfiguration problem consists of a formula together with two satisfying
assignments. The satisfying assignments that we will construct will both set the variables $x,y,z,t$ to those values.  

\begin{figure}
\begin{center}
\includegraphics[page=2,scale=0.4]{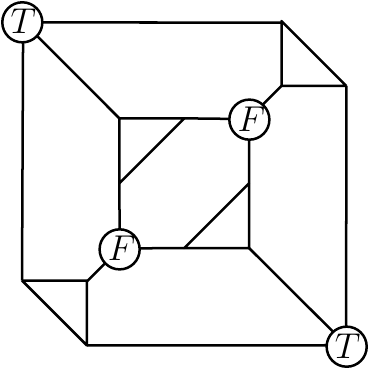}
\end{center}
\caption{\label{fig:rigid}A set of monotone NAE clauses together with a satisfying assignment, no variable of which can be flipped.}
\end{figure}

In order to encode the orientation of the edges of the constraint graph, we define one Boolean variable for each edge $x$. For a given node of the constraint graph, we will denote by $\xin$ the literal that is true whenever the edge $x$ is oriented towards the node, and by $\xout$ the literal that is true whenever $x$ points outwards. (Note that this notation is relative to a given node.)

We now consider the AND nodes in the input constraint graph. For each such node, we include a new set of NAE clauses in the instance of the reconfiguration problem. Since the clauses must be monotone, we have to allow for different cases, depending which orientation is encoded by each of the edge variables. Let us consider an AND node with one blue edge $x$ and two red edges $y$ and $z$.

Let $\xin$ be the literal that is true whenever the edge $x$ is oriented towards the node. 
We can safely assume that this literal is positive, that is, setting the {\em variable} to true means that the edge is directed towards the node. This is without loss of generality, because in an instance of NAE 3-SAT, all clauses can be safely replaced by the same clause with all literals negated. Hence in the case where $\xout$ is the positive literal, we can use the same sets of clauses with all literals negated.

Given this, it remains to take into account all situations involving negations of the literals $\yin$ and $\zin$ in order to have only monotone clauses. First suppose that the literals $\yin$ and $\zin$ are positive, that is, the variables for $y$ and $z$ correspond to the incoming orientation. Then we include the following clauses:
\begin{equation}
    \label{gadget:and1}
(\yin \vee \xin \vee F) \wedge (\zin \vee \xin \vee F),
\end{equation}
where $F$ denotes two variables that has been set to false using the previous construction.
In the case where the variable associated with edge $y$ corresponds to the outgoing orientation and $z$ corresponds to the incoming orientation, we include the following three clauses:
\begin{equation}
    \label{gadget:and2}
(\yout \vee T \vee t) \wedge (\zin \vee F \vee \xin) \wedge (t \vee F \vee \xin),
\end{equation}
where $T,F$ denote three distinct variables that have been set to true and false, respectively, using the previous construction, and $t$ is an additional variable that does not appear anywhere else.
Finally, if both variables for $y$ and $z$ correspond to the outgoing orientation, we include the following four clauses:
\begin{equation}
    \label{gadget:and3}
(\yout \vee T \vee t) \wedge (\zout \vee T \vee t') \wedge (t \vee F \vee \xin) \wedge (t' \vee F \vee \xin),  
\end{equation}
where again $T$ and $F$ denote distinct variables that have been set to true and false, respectively, and $t,t'$ are new variables.
The clauses are illustrated in Figure~\ref{fig:ANDOR}.
In all three cases, the set of clauses force both $\yin$ and $\zin$ to be true (that is, $\xout$ and $\yout$ to be false) whenever $\xin$ is false.
On the other hand, if $\xin$ is true, then the variables associated with the edges $y$ and $z$ can take any value.
Hence this properly encodes the semantics of an AND node in NCL.

We now describe how to implement the OR nodes of the constraint graph as collections of monotone NAE clauses.
We consider an OR node with three incident blue edges denoted by $x,y,z$. Recall that the only forbidden orientation is the one in which all three edges are outgoing.
For the same reason as before, we can safely assume that the variable associated with $x$ encodes the incoming orientation, that is, the literal $\xin$ is positive.
In the case where the two variables associated with $y$ and $z$ correspond to the outgoing orientation, that is, the literals $\yout$ and $\zout$ are positive, then
we include the following three clauses:
\begin{equation}
    \label{gadget:or1}
(\yout \vee T \vee t) \wedge (\zout \vee T \vee t') \wedge (t \vee t' \vee \xin).
\end{equation}
If the literals $\yin$ and $\zin$ are both positive, then we include the following four clauses:
\begin{equation}
      \label{gadget:or2}
(\xin \vee F \vee t'') \wedge (\yin \vee F \vee t') \wedge (\zin \vee F \vee t) \wedge (t \vee t' \vee t'').
\end{equation}
Finally, if $\yin$ and $\zout$ are positive, we include:
\begin{equation}
      \label{gadget:or3}
(\zout \vee T \vee t') \wedge (\yin \vee t' \vee \xin).
\end{equation}
Again, the symbols $T,F$ denote variables whose values have been set to true or false, and $t,t',t''$ are new variables.
The constructions are illustrated in Figure~\ref{fig:ANDOR}.
It can be checked that in all three cases, the only forbidden assignment is the
one in which all three edges are outgoing. Hence this correctly encodes the semantics of an OR node in NCL.

Finally, we need to provide the two satisfying assignments for the NAE SAT reconfiguration instance. Those assignments are constructed from the initial
configurations $C_1$ and $C_2$ of the input C2C instance. From this, we directly infer the values of the variables associated with each edge of the constraint graph.
There are also two types of additional variables. Those whose values are fixed by the clauses in \eqref{gadget:rigid} have the same value in both assignments. Those denoted
by $t,t',t''$ in the clauses encoding the nodes of the constraint graph get an arbitrary value such that all clauses are satisfied, which is always possible by construction.

It can further be checked that for all sets of clauses encoding the nodes of the constraint graph have, the variable-clause incidence graph is a tree. The incidence graph
of the clauses in \eqref{gadget:rigid} is planar. Altogether, the variable-clause incidence graph of the output NAE SAT instance is also planar.

We now prove the correctness of the construction. First, suppose that there exists a reconfiguration sequence for the output NAE SAT instance. By projecting each assignment in this sequence on the edge variables, we obtain a reconfiguration sequence for the constraint graph. From the validity of the construction, this sequence must be valid for the C2C problem as well.

Suppose now that there exists a reconfiguration sequence for the input C2C instance. We show that this sequence can be mapped to a reconfiguration sequence for the NAE SAT instance.
We only need to check that any valid flip of an edge in the constraint graph can be mapped to a sequence of flips of the variables involved in the constructions. The edge flips naturally maps
to flips of variables encoding the edge orientation. However, to maintain a satisfying assignment at every step, we also need to flip some of the new variables of the form $t,t',t''$ in the
clauses of the form \eqref{gadget:and2}-\eqref{gadget:or3}.

For the constructions \eqref{gadget:and2} and \eqref{gadget:and3}, we observe that $t$ and $t'$ can always be set to false, except when $\xin$ is false. In the latter case, the only edge variable that we can flip is $\xin$. We can always flip $\xin$ to true while keeping $t$ and $t'$ set to true. We can then then flip $t$ and $t'$ to false afterwards. symmetrically, flipping $\xin$ back to false can always be done by first flipping $t$ and $t'$ to true.

Similarly, for the constructions \eqref{gadget:or1}-\eqref{gadget:or3}, one can check that we can flip any edge variable provided the orientation remains legal, possibly by flipping some of the variables $t,t',t''$ in intermediate steps. Let us detail, for instance, case \eqref{gadget:or2}, in which  setting some of $\xin$, $\yin$, or $\zin$ to false forces one of $t,t',t''$ to be set to true.
First consider flips involving assignments in which not all three variables $\xin, \yin, \zin$ are true simultaneously. For these assignments, we take the convention that $t,t',t''$ is set to false unless it is forced to be true. Suppose without loss of generality that $\xin$ flips from false to true. It suffices to first flip $\xin$, then flip $t$ from true to false. Now when $\xin, \yin$, and $\zin$ are all true, any assignment of the three variables $t,t',t''$ is satisfying. Hence flips from or to this situation can be handled as well.

By intertwining, for each edge flip, the sequences of variable flips in all gadgets
containing the variable for this edge, we can map the reconfiguration sequence of the C2C
problem to a reconfiguration sequence between the two given satisfying assignments.
This concludes the proof.
\end{proof}

\section{Subset Sum Reconfiguration}
\label{sec:subset}

We now consider the reconfiguration problem for the well-known subset sum problem.

\begin{definition}[Subset Sum Problem]
Given an integer $x$ and a set of integers $S = \{a_1, a_2, \dots, a_n\}$, does there exist a subset $A\subseteq [n]$ such that $\sum_{i \in A}{a_i} = x$?
\end{definition}

If we restrict our reconfiguration steps to involve only a single element of $S$, the reconfiguration problem is trivial, as no single such move can maintain the same sum. We therefore consider more general reconfiguration steps. We say that a set of integers $A_1$ can be \emph{$k$-move reconfigured} into a second set of integers $A_2$ whenever the symmetric difference of $A_1$ and $A_2$ has cardinality at most $k$. 

\begin{definition}[$k$-move Subset Sum Reconfiguration Problem]
  Given two solutions $A_1$ and $A_2$ to an instance of the subset sum problem, can $A_2$ be obtained by repeated $k$-move reconfiguration, beginning with $A_1$, so that all intermediate subsets are also solutions?
\end{definition}

The problem remains trivial for $k=2$, since any removed element must be replaced by itself. For $k=3$, we prove the following theorem.

\begin{theorem}
\label{thm:unary-3-move}
The 3-move subset sum reconfiguration problem is strongly \PSPACE{}-complete.
\end{theorem}

The problem is \emph{strongly} \PSPACE{}-complete, meaning that it remains \PSPACE{}-complete when the input integer set is given in unary.
The corresponding instances of the subset sum problem can be solved in polynomial time using dynamic programming. 
This is another example of a reconfiguration problem that is \PSPACE{}-complete, despite the underlying decision problem lying in \P{}.
We first note that the problem is contained in \PSPACE{}. 

\begin{lemma}
For every $k \in \mathbb{N}$, the $k$-move subset sum reconfiguration problem is in \PSPACE{}.
\end{lemma}
\begin{proof}
The proof is a slight modification of a proof for a variation of the problem due to~\cite{Ito-2014a}.
For an instance with $|S| = n$, there are $O(n^k)$ other subsets reachable by a $k$-move reconfiguration, since each such move can be specified by the set of items in the symmetric difference of the two subsets.
So all adjacent subsets in the reconfiguration graph can be enumerated in polynomial time.

Then the $k$-move subset sum reconfiguration problem is in \NPSPACE{} by the following algorithm: in the reconfiguration graph, repeatedly move between subsets by non-deterministically selecting a neighbor in polynomial time (and space).
Since $\NPSPACE{} = \PSPACE{}$~\cite{Savitch-1970a}, the problem is also in \PSPACE{}.
\end{proof}

As for hardness, the reduction is done in two steps.
First, from the Sliding Tokens problem to the Exact Cover reconfiguration problem (Lemma~\ref{lem:exact-cover}), then to the 3-move Subset Sum reconfiguration problem (Theorem~\ref{thm:unary-3-move}).

\begin{definition}[Token Slide Reconfiguration]
Given two independent sets $I_1$, $I_2$ of a graph $G = (V, E)$, $I_1$ can be \emph{reconfigured} into $I_2$ via a \emph{token slide} provided $(I_1 - I_2) \cup (I_2 - I_1) = \{v_1, v_2\}$ and $\{v_1, v_2\} \in E$.
\end{definition}

Observe that a token slide corresponds to changing the selection of a vertex $v_1 \in I_1$ to a neighboring vertex $v_2 \in I_2$, possible exactly when $v_1$ is the only vertex in $I_1$ among $v_1$, $v_2$, and their neighbors.

\begin{definition}[Sliding Tokens Problem]
Given two independent sets $I_1$, $I_2$, can $I_1$ be reconfigured into $I_2$ via repeated token slides?
\end{definition}

An \emph{exact cover} is a set cover that covers every element exactly once.

\begin{definition}[Exact Cover Split and Merge Reconfiguration]
Given a set $\mathcal{S}$ of subsets of a set $U$, and two exact covers $C_1, C_2 \subseteq \mathcal{S}$, $C_1$ can be \emph{reconfigured} into $C_2$ via a \emph{split} (and $C_2$ can be reconfigured into $C_1$ via a \emph{merge}) provided that there exist $S_1, S_2, S_3 \subseteq \mathcal{S}$ with $C_1 - C_2 = S_1$ and $C_2 - C_1 = \{S_2, S_3\}$.
\end{definition}

Since $C_1$, $C_2$ are exact covers, $S_1 = S_2 \cup S_3$ and $S_2 \cap S_3 = \varnothing$. 

\begin{definition}[Exact Cover Reconfiguration Problem]
Given a set $\mathcal{S}$ of subsets of a set $U$, can $C_1$ be reconfigured into $C_2$ via repeated splits and merges?
\end{definition}

Recall that a set $\mathcal{S}$ of subsets of a set $U$ can be considered as a hypergraph $G = (U, \mathcal{S})$, where each element of $U$ is a vertex and each element of $\mathcal{S}$ is a hyperedge. We say that a hypergraph is {\em $k$-colorable} whenever we can assign one of $k$ colors to each vertex such that no two vertices in a hyperedge have the same color.

\begin{lemma}
\label{lem:exact-cover}
The exact cover reconfiguration problem is \PSPACE{}-hard for instances that are 23-colorable hypergraphs.
\end{lemma}
\begin{proof}
The proof of Theorem~23 of~\cite{Hearn-2005a} establishes that the sliding tokens problem is \PSPACE-hard on 3-regular graphs (see Section 3.2 of~\cite{BC09} for further discussion). A trivial modification of the proof suffices to prove that a \emph{labeled} variant of the sliding tokens problem, where each token has a unique label, is also \PSPACE-hard. The reduction is from this variant. The following describes an input instance of the labeled sliding tokens problem:
\begin{itemize}
\item $G = (V, E)$, a 3-regular graph.
\item $T$, a set of labeled tokens.
\item $p_1 : T \rightarrow V$, a function mapping each labeled token to a vertex placement in the starting configuration.
\item $p_2 : T \rightarrow V$, a function mapping each labeled token to a vertex placement in the ending configuration.
\end{itemize} 
Also, $I_1 = \{ p_1(t) : t \in T \}$ and $I_2 = \{ p_2(t) : t \in T\}$ are independent sets of size $|T| \leq |V|$.

\paragraph{Output $U$ and $\mathcal{S}$.}
The output exact cover instance has a set $U$ consisting of two types of elements: \emph{vertices} $v_1, v_2, \dots, v_{|V|}$ and \emph{tokens} $t_1, t_2, \dots, t_{|T|}$.
That is, $U = \{v_1, v_2, \dots, v_{|V|} \} \cup \{t_1, t_2, \dots, t_{|T|} \}$.

For each pair of adjacent vertices $v_i, v_j \in V$, the set consisting of these two vertices and their neighbors is called a \emph{slide set}, denoted $S_{i, j}$.
The output set $\mathcal{S}$ of subsets of $U$ contains the following subsets for every pair of adjacent vertices $v_i$, $v_j$ and token $t_k$:
\begin{itemize}
\item All subsets of $S_{i, j} - \{v_i\}$ and $S_{i, j} - \{v_j\}$.
\item $\{v_i, t_k\}$ and $\{v_j, t_k\}$.
\item $S_{i, j} \cup \{t_k\}$.
\end{itemize}

\paragraph{Output $C_1$ and $C_2$.}
The starting configuration $C_1$ is the union of $\{ \{v_i\} : v_i \in V - I_1 \}$ and, for every $v_i \in I_1$, a set $\{v_i, t_k\}$ with a distinct $t_k$.
Similarly, the ending configuration $C_2$ is the union of $\{ \{v_i\} : v_i \in V - I_2 \}$ and, for every $v_i \in I_2$, a set $\{v_i, t_k\}$ with a distinct $t_k$.

\paragraph{23-colorability of $(U, \mathcal{S})$.}
Since $G$ is 3-regular, $G^3$ has degree at most~21. 
So $G$ can be 22-colored such that no two vertices of distance at most 3 (i.e. in a common slide set) have the same color.
Such a coloring ensures that no pair of vertices in a common set in $\mathcal{S}$ share a color.
Coloring the tokens in $T$ a distinct (23rd) color then gives a coloring of $U$ such that no pair of elements of a common set share the same color. 

\paragraph{High-level idea.}
The subsets containing exactly one vertex and token (e.g., $\{v_i, t_k\}$) represent the presence of the token $t_k$ on vertex $v_i$.
Subsets consisting of a slide set and token (e.g., $S_{i, j} \cup \{t_k\}$) represent the presence of a ``mid-slide'' token between $v_i$ and $v_j$.

Sliding a token $t_k$ from $v_i$ to $v_j$ is simulated by first merging $\{v_i, t_k\}$ and $S_{i, j} - \{v_i\}$ into $S_{i, j} \cup \{t_k\}$, and then splitting this set into $S_{i, j} - \{v_j\}$ and $\{v_j, t_k\}$.
This sequence enforces the absence of tokens on neighbors of $v_i$ and $v_j$, and the presence of a token on $v_i$ or $v_j$, but not both.
Before a merge-split sequence, additional splits and merges of token-less sets may be needed to obtain $S_{i, j} - \{v_i\}$.

\newcommand{\redfunce}{\ensuremath{f_{\mathrm{red}}}}
\newcommand{\redfunc}[1]{\ensuremath{f_{\mathrm{red}}(#1)}}

\paragraph{Bijection between configurations.}
Call a configuration $C$ of the output Exact Cover Reconfiguration instance \emph{maximally split} if every $C$ in $C$ contains exactly one vertex and up to one token.
The following defines a function \redfunce{} from token arrangements to maximally split covers:
\begin{itemize}
\item Each token-less vertex corresponds to a set $\{v_i\}$ in the cover.
\item Each token $t_k$ placed at $v_i$ corresponds to a set $\{v_j, t_k\}$ in the cover.
\end{itemize}
Notice that \redfunce{} is a bijection and $\redfunc{p_1} = C_1$, $\redfunc{p_2} = C_2$.

\paragraph{Reduction structure.}
The remainder of the proof is devoted to proving the following claim: a token arrangement $p'$ is reachable from a token arrangement $p$ if and only if $\redfunc{p'}$ is reachable from $\redfunc{p}$ via splits and merges.

Both directions are proved inductively.
That is, we consider only ``adjacent'' configurations.
We also assume that the starting token arrangement $p : T \rightarrow V$ has $\{ p(t) : t \in T\}$ independent.

\paragraph{Sliding tokens reachability $\Rightarrow$ exact cover reachability.} 
Let $p$ be a token arrangement that can be reconfigured into $p'$ via a token slide from $v_i$ to $v_j$.
Then $\redfunc{p'}$ can be reached from $\redfunc{p}$ via the following sequence of merges and splits:
\begin{enumerate}
\item Repeatedly merge token-less vertex sets to form $S_{i, j} - \{v_i\}$.
\item Merge $S_{i, j} - \{v_i\}$ and $\{v_i, t_k\}$ into $S_{i, j} \cup \{t_k\}$.
\item Split $S_{i, j} \cup \{t_k\}$ into $S_{i, j} - \{v_j\}$ and $\{v_j, t_k\}$.
\item Repeatedly split the token-less vertex set $S_{i, j} - \{v_j\}$ into single vertex sets.
\end{enumerate}

\newcommand{\sploot}[1]{\ensuremath{\mathrm{sploot}(#1)}}

\paragraph{Exact cover reachability $\Rightarrow$ sliding tokens reachability.} 
For each exact cover configuration $C$ in the output instance, at least one maximally split configuration is reachable from $C$ via a sequence of splits.
Call the set of all such configurations the \emph{sploot set} of $C$, denoted $\sploot{C}$.

\newcommand{\cinter}{\ensuremath{C_{\mathrm{inter}}}}

Let $C$, $C'$ be maximally split configurations such that $C$ can be reconfigured into $C'$ and \cinter{} is the first configuration reached such that $\sploot{\cinter} \neq \{C\}$.
By induction, assume $C' \in \sploot{\cinter}$.

Since splits and token-less merges do not add elements to a sploot set, \cinter{} is obtained by merging two sets, one of which contains a token.
Since the only token-containing sets that can be merged are those of the form $\{v_i, t_k\}$, \cinter{} is obtained by merging $\{v_i, t_k\}$ and $S_{i, j} - \{v_i, t_k\}$ to obtain $S_{i, j} \cup \{t_k\}$ for some $v_i$, $v_j$, and $t_k$. 
Notice that it may be the case that $S_{i, j} = S_{i', j'}$ for other pairs $i', j'$.
 
% Start new 
Such a merge allows two kinds of splits:
\begin{itemize}
\item Splitting $S_{i, j}$ into $S_{i, j} - \{v_i, t_k\}$ (to obtain the previous configuration, with sploot set $\{C\}$).
\item Splitting $S_{i, j}$ into $S_{i', j'} - \{v_j', t_k\}$, where $S_{i, j} = S_{i', j'}$ (to obtain a new configuration with sploot set $\{C'\}$, where $C'$ is identical to $C$, except that $C'$ contains $\{v_j', t_k\}, \{v_i'\}$ instead of $\{v_i, t_k\}, \{v_j\}$).
\end{itemize}

% Intuitively: 
% -S_{i, j} \cup S_{i', j'} contains only one token (at v_i) since it equals S_{i, j}.
% -So can slide along any edge between vertices in {v_i, v_j, v_i', v_j'}.
% -There is at least one edge between a vertex of {v_i, v_j} and a vertex of {v_i', v_j'},
%  since otherwise S_{i, j} does not include v_i' or v_j' and thus is not equal to S_{i', j'}. 
% -So can slide from v_i to v_j' via at most 3 slides.

Since $S_{i, j} - \{v_i\}, \{v_j, t_k\} \in C$, the token arrangement $p$ with $\redfunc{p} = C$ has no tokens on vertices in $S_{i, j}$ except for token $t_k$ on $v_i$.
Since $S_{i, j} \cup S_{i', j'} = S_{i, j}$ contains all neighbors of $v_i$, $v_j$, $v_i'$, $v_j'$, the token arrangement obtained by moving the location of $t_k$ in $p$ from $v_i$ to $v_j$, $v_i'$, or $v_j'$ is an independent set. 

So all that remains is to prove that there are a sequence of slides moving $t_k$ from $v_i$ to $v_j'$ via vertices in $\{v_i, v_j, v_i', v_j'\}$.
Since $S_{i, j} = S_{i', j'}$, $v_i', v_j' \in S_{i, j}$ and so either $v_i \in \{v_i', v_j'\}$, or there is an edge $\{v_i, v_i'\}$ or $\{v_i, v_j'\} \in E$.
So $t_k$ can slide from $v_i$ to either $v_i'$ or $v_j'$ (via~0 or~1 slides), and then from $v_i'$ or $v_j'$ to $v_j'$ (via~0 or~1 slides).
\end{proof}

We are now ready to prove the main result of this section.

\begingroup
\def\thetheorem{\ref{thm:unary-3-move}}
\begin{theorem}
The 3-move subset sum reconfiguration problem is strongly~\PSPACE{}-complete.
\end{theorem}
\addtocounter{theorem}{-1}
\endgroup

\begin{proof}
The reduction is from the Exact Cover reconfiguration problem for instances that are 23-colorable induced hypergraphs, proved \PSPACE-hard by Lemma~\ref{lem:exact-cover}. 
Observe that every 3-move subset sum reconfiguration is either a \emph{merge}, where $a_i$ and $a_j$ are replaced by $a_i + a_j$, or a \emph{split}, where $a_i + a_j$ is replaced by $a_i$ and $a_j$.
Each set split or merge will correspond to a 3-move split or merge, respectively, in the output instance.

\paragraph{Output numbers and sum.}
A function $f : U \rightarrow \mathbb{N}$ maps each element of the universe $U$ of the input exact cover reconfiguration problem to a positive integer, and the numbers in the output 3-move subset sum reconfiguration instance are $\{ \sum_{a \in S}f(a) : S \in \mathcal{S}\}$ and the output target sum is $\sum_{a \in U}f(a)$.

Elements of $U$ are partitioned according to their colors $1, 2, \dots, 23$ and (arbitrarily) labeled $a_1, a_2, \dots, a_{|U|}$.
The function $f$ maps a color-$j$ element $a_i$ to $i\cdot 2^{100j \lceil \log_2(|U|) \rceil}$.
In binary, this mapping consists of the binary encoding of $i$ followed by by $100j \lceil \log_2(|U|) \rceil$ zeros.

\paragraph{Output size.}
The output instance consists of $|\mathcal{S}|$ numbers, each between 0 and $|U| \cdot 2^{100 \cdot 23 \lceil \log_2(|U|) \rceil} = O(|U|^2)$.
So the output sum is $O(|U|^3)$. 
Thus the output instance, encoded in unary, has length $O(|\mathcal{S}||U|^2 + |U|^3)$, i.e. polynomial in the input instance.

\paragraph{Correctness.}
A reconfiguration in both the exact cover and 3-move subset sum problems involves splitting or merging elements.
Thus it suffices to prove that the function $f$ yields a one-to-one mapping $g : \mathcal{S} \rightarrow \mathbb{N}$ given by $g(S) = \sum_{a \in S}f(a)$.

Recall that the function $f$ maps each element $a_i \in U$ to a value based upon the color of $a_i$.
The sums of the outputs of $f$ for all elements of all colors $1$ to $j-1$ is at most $2^{100(j-1) \lceil \log_2(|U|) \rceil} \cdot |U|^2 \leq 2^{(100j - 98) \lceil \log_2(|U|) \rceil}$
while the output of $f$ for any element of any color $j$ or larger is at least $2^{100j \lceil \log_2(|U|) \rceil} \geq 2^{98} \cdot 2^{(100j - 98) \lceil \log_2(|U|) \rceil}$. 

Thus if a pair of sets $S_1, S_2 \subseteq \mathcal{S}$ have $S_1 \neq S_2$, then their color-$j$ elements differ, this difference cannot be made up by adding or removing elements of colors 1 to $j-1$ (values too small) or colors $j+1$ to $23$ (values too large).
Thus if $S_1 \neq S_2$, then $g(S_1) \neq g(S_2)$.
\end{proof}

\section{Reconfiguration Problems and Paths in Hypercubes}
\label{sec:paths}

The $n$-hypercube is the graph with vertex set $\{0,1 \}^n$ such that two vertices are adjacent whenever their coordinates differ by exactly one component. In this section, we consider the following abstraction of reconfiguration problems involving subsets.

\begin{definition}[Constrained Hypercube Path] Given two vertices $s,t$ of the $n$-hypercube, both contained in a polytope $P:=\{x\in \mathbb{R}^n : Ax\leq b\}$  for some $A=(a_{ij})\in\mathbb{Z}^{d\times n}$ and $b\in\mathbb{Z}^d$, does there exist a path from $s$ to $t$ in the hypercube, all vertices of which lie in $P$?
\end{definition}

The constrained hypercube path problem can be seen as a reconfiguration analogue of the {\em 0-1 integer linear programming} (0-1 ILP) satisfiability problem, which simply asks for the existence of a 0-1 point in the inside \P{}, and is a standard \NP{}-complete problem from Karp's list. (Note that this problem is distinct from the 0-1 ILP Reconfiguration problem defined in Ito et al.~\cite{Ito-2011a}: in the latter, a solution must optimize some objective function, while we are only concerned with satisfiability.)

The subset sum problem is the question of the existence of a 0-1 point in a polytope consisting of a subspace of dimension $n-1$, hence defined by two linear constraints with the same coefficients. 
Similarly, the knapsack (decision) problem involves exactly two linear constraints, and the Knapsack reconfiguration problem can be cast as a special case of the constrained hypercube path problem where $d=2$. The definitions are as follows.

\begin{definition}[Knapsack Problem]
Given integers $\ell$ and $u$ and two sets of integers $S = \{a_1, a_2, \dots, a_n\}$ and $W = \{w_1, w_2, \ldots , w_n\}$, does there exist a subset $A\subseteq [n]$ such that $\sum_{i \in A}{a_i} \geq \ell$ and $\sum_{i\in A}{w_i}\leq u$?
\end{definition}

\begin{definition}[Knapsack Reconfiguration Problem]
  Given two solutions $A_1$ and $A_2$ to an instance of the knapsack problem, can $A_2$ be obtained by repeated $1$-move reconfiguration, beginning with $A_1$, so that all intermediate subsets are also solutions?
\end{definition}

Demaine and Ito considered the knapsack reconfiguration problem in the case where $S=W$~\cite{Ito-2014a}. They proved that the problem was \NP{}-hard, and gave an approximation algorithm for finding a reconfiguration sequence in which the intermediate steps satisfy one of the constraints only up to some multiplicative factor. Whether the knapsack reconfiguration problem is \PSPACE{}-complete is a tantalizing open question.
Characterizing the complexity of the knapsack reconfiguration problem implies understanding the complexity of the constrained hypercube path problem for bounded values of $d$. We do not settle the former question, but provide an answer to the latter.
The proof of Theorem~\ref{thm:paths} uses a reduction from a variant of the exact cover reconfiguration problem from the proof of Theorem~\ref{thm:unary-3-move} where more-than-2-way merges and splits are also permitted:

\begin{definition}[Partition and Union Reconfiguration]
Given a set $\mathcal{S}$ of subsets of a set $U$, and two exact covers $C_1, C_2 \subseteq \mathcal{S}$, $C_1$ can be \emph{reconfigured} into $C_2$ via a \emph{partition} (and $C_2$ can be reconfigured into $C_1$ via a \emph{union}) provided that there exist $S_1, S_2, S_3, \dots, S_k \subseteq \mathcal{S}$ with $C_1 - C_2 = S_1$ and $C_2 - C_1 = \{S_2, S_3, \dots, S_k\}$.
\end{definition}

\begin{definition}[Exact Cover Many-Way Reconfiguration Problem]
Given a set $\mathcal{S}$ of subsets of a set $U$, can $C_1$ be reconfigured into $C_2$ via repeated partitions and unions?
\end{definition}

The reduction given in the proof of Lemma~\ref{lem:exact-cover} also proves that this variant is hard. 
This can be seen by observing that any set that can be formed ``repeatedly merging'' or set of sets that can be formed by ``repeatedly splitting'' can also be formed by a single union or partition.

\begin{corollary}
\label{cor:exact-cover-manyway}
The exact cover many-way reconfiguration problem is \PSPACE{}-hard for instances that are 23-colorable hypergraphs.
\end{corollary}

\begin{theorem}
\label{thm:paths}
The Constrained Hypercube Path problem is \PSPACE{}-complete, even when $d=O(1)$.
\end{theorem}

\begin{proof}
The constrained hypercube path problem is equivalent to the generalization of the knapsack reconfiguration problem, where each integer is instead a multi-dimensional tuple, and the sum of the elements in each dimension must lie in a specified range.

The reduction is from the exact cover many-way reconfiguration problem, and is a modification of the reduction given in the proof of Theorem~\ref{thm:unary-3-move}.
Instead of mapping color-$j$ elements to values in the range $2^{100j \lceil \log_2(|U|) \rceil}$ to $|U| \cdot 2^{100j \lceil \log_2(|U|) \rceil}$, a magnitude unique to $j$, color-$j$ elements are mapped to a set of three dimensions unique to $j$.

The reconfigurations permitted in the Exact Cover Many-Way Reconfiguration Problem are simulated by sequences of ``1-move'' reconfigurations (adding or removing an element) permitted in the Knapsack Reconfiguration Problem.
An additional dimension limits the use of ``key'' elements that ``unlock'' small portions of the cover, enabling reconfiguration.

\newcommand{\cdimpv}[1]{\ensuremath{\mathrm{dim}^{\mathrm{val}+}_{#1}}}
\newcommand{\cdimnv}[1]{\ensuremath{\mathrm{dim}^{\mathrm{val}-}_{#1}}}
\newcommand{\cdimc}[1]{\ensuremath{\mathrm{dim}^\mathrm{cnt}_{#1}}}
\newcommand{\kdim}{\ensuremath{\mathrm{dim}^\mathrm{key}}}

\paragraph{Dimensions.}
There are $23\cdot3 + 1 = d$ dimensions:
\begin{itemize}
\item Three \emph{color} dimensions for each color $j$: \emph{positive} and \emph{negative} \emph{value} dimensions denoted \cdimpv{j} and \cdimnv{j}, respectively, and a \emph{count} dimension denoted \cdimc{j}.
\item A \emph{key} dimension denoted \kdim. % Used to allow up to 1 key value.
\end{itemize}

\newcommand{\colorfunce}{\ensuremath{f_{\mathrm{col}}}}
\newcommand{\colorfunc}[1]{\ensuremath{f_{\mathrm{col}}(#1)}}

\paragraph{Mapping colors and elements.}
The reduction uses two functions.
The first, $\colorfunce : U \rightarrow [23]$, maps the elements of $U$ to their respective colors in a 23-coloring of $U$ according to the hypergraph $\mathcal{S}$.

\newcommand{\unifunce}{\ensuremath{f_{\mathrm{uni}}}}
\newcommand{\unifunc}[1]{\ensuremath{f_{\mathrm{uni}}(#1)}}

The second, $\unifunce : U \rightarrow \mathbb{N}^d$ maps elements of the universe $U$ of the input exact cover reconfiguration problem to a $d$-dimensional tuple.
The function \unifunce{} maps a color-$j$ element $a_i$ to a $d$-dimensional vector $\vec{v}$ that is 0-valued in all dimensions except three:
\begin{itemize}
\item \cdimpv{j}, where $\vec{v}$ has value $i$.
\item \cdimnv{j}, where $\vec{v}$ has value $|U|+1-i$.
\item \cdimc{j}, where $\vec{v}$ has value 1.
\end{itemize}

\newcommand{\tupfunce}{\ensuremath{f_{\mathrm{tup}}}}
\newcommand{\tupfunc}[1]{\ensuremath{f_{\mathrm{tup}}(#1)}}
\newcommand{\keyfunce}{\ensuremath{f_{\mathrm{key}}}}
\newcommand{\keyfunc}[1]{\ensuremath{f_{\mathrm{key}}(#1)}}

\paragraph{Output tuples and sum range.}
There are two kinds of output tuples: \emph{set tuples} and \emph{key tuples}. 
For each set $S \in \mathcal{S}$, there is one output set tuple $\tupfunc{S} = \sum_{a \in S}{\unifunc{a}}$.

For each set $S \in \mathcal{S}$, there is also a key tuple with the same values as $\tupfunc{S}$ in all dimensions $\cdimpv{j}$ and $\cdimnv{j}$, 0-valued in all color count dimensions, and value~1 in dimension \kdim{}.

For each color $j$, let $A_j = \{ a \in U : \colorfunc{a} = j \}$.
For each \cdimpv{j} and \cdimnv{j}, the output target sum has minimum values $\sum_{a_i \in A_j}{i}$ and $\sum_{a_i \in A_j}{|U|-i+1}$, respectively.
For each \cdimc{j}, the sum has maximum value~$|A_j|$.
For \kdim{}, the sum has maximum value~1.

\paragraph{Output size.}
The output instance consists of $2|\mathcal{S}|$ tuples, with the value in each dimension of each tuple between 0 and $|U|$.
Also, the sum range in each dimension consists of two values between 0 and $|U|^2$.
So the output instance, encoded in unary, has length $O(|\mathcal{S}||U| + |U|^2)$, i.e. polynomial in the input instance.

\newcommand{\figfunce}{\ensuremath{f_{\mathrm{fig}}}}
\newcommand{\figfunc}[1]{\ensuremath{f_{\mathrm{fig}}(#1)}}

\paragraph{Bijection between configurations.}
Let $\mathcal{T}$ be the set of set tuples in the output generalized Knapsack Reconfiguration Problem instance.
Recall that each set $S \in \mathcal{S}$ defines an output set tuple via the function \tupfunce{}.
Define the function $\figfunce : \mathcal{P}(\mathcal{S}) \rightarrow \mathcal{P}(\mathcal{T})$ from subsets of $\mathcal{S}$ to subsets of $\mathcal{T}$ to be the mapping $\figfunc{C} = \{ \tupfunc{S} : S \in C \}$.

Since \tupfunce{} is a bijection between elements of $\mathcal{S}$ and $\mathcal{T}$, and \figfunce{} maps each subset of $\mathcal{S}$ to the corresponding subset of $\mathcal{T}$ via \tupfunce{}, \figfunce{} is a bijection between $\mathcal{P}(\mathcal{S})$ and $\mathcal{P}(\mathcal{T})$.

\paragraph{Reduction structure.}
The remainder of the proof is devoted to proving the following claim: an exact cover $C_2$ is reachable from $C_1$ via partitions and unions if and only if $\figfunc{C_2}$ is reachable from $\figfunc{C_1}$ via 1-move reconfigurations.

Both directions of the claim are proved inductively.
That is, we consider only ``adjacent'' configurations: exact covers that differ by one partition or union, and key-tuple-less tuple sets that can be reconfigured into one another without visiting intermediate key-tuple-less tuple sets.
Before proving each direction, we make a few observations about the output instance. 

\paragraph{Observation 1: at most one key tuple.}
Since the maximum value in \kdim{} is 1, each key tuple has value~1 in \kdim{}, and set tuples have value 0 in \kdim{}, at most one key tuple is present in any valid configuration.

\paragraph{Observation 2: sums of key-tuple-less configurations are tight.} 
For any exact set cover $C \subseteq \mathcal{S}$ of $U$, the sum of the elements in $\figfunc{C}$ has the minimum value in \cdimpv{j}, \cdimnv{j} and maximum value in \cdimc{j} for all $j$.

\paragraph{Exact cover reachability $\Rightarrow$ knapsack reachability.} 
We consider the case of reconfiguring an exact cover into another via a partition; the case of a union is symmetric. 
Suppose there exist two exact covers $C_1$ and $C_2$ with $C_1 - C_2 = S_1$ and $C_2 - C_1 = \{S_2, \dots, S_k\}$.
Then $\figfunc{C_1}$ can be reconfigured into $\figfunc{C_2}$ via the following moves:
\begin{enumerate}
\item Add the key tuple $\keyfunc{S_1}$.
\item Remove the set tuple $\tupfunc{S_1}$.
\item Add the set tuples $\tupfunc{S_2}, \dots, \tupfunc{S_k}$.
\item Remove the key tuple $\keyfunc{S_1}$.
\end{enumerate}

\paragraph{Knapsack reachability $\Rightarrow$ exact cover reachability.}
Since $\figfunce$ is a bijection, any pair of key-tuple-less subsets of $\mathcal{T}$ can be written as $\figfunc{C_1}$, $\figfunc{C_2}$ where $C_1, C_2 \subseteq \mathcal{S}$.

Suppose $\figfunc{C_1}$ can be reconfigured into $\figfunc{C_2}$.
By previous observations, reconfiguring $\figfunc{C_1}$ is only possible via adding a key tuple, and only one key tuple can ever be present.
So the move sequence must have the following form:
\begin{enumerate}
\item Add a key tuple $\keyfunc{S_1}$ for some $S_1 \in \mathcal{S}$.
\item Add and remove set tuples.
\item Remove $\keyfunc{S_1}$.
\end{enumerate}

Upon adding a key tuple $\keyfunc{S_1}$ with $a_i \in S_1$ and $\colorfunc{a_i} = j$, the sum values in dimensions $\cdimpv{j}$ and $\cdimnv{j}$ are increased above the minimum values by $i$ and $|U|-i+1$, respectively.
For each such $a_i$, removing a set tuple whose set contains any other element $a_i'$ with $\colorfunc{a_i'} = j$ is not possible, since then: 
\begin{itemize}
\item $i > i'$ and the dimension $\cdimpv{j}$ sum falls below the minimum, or
\item $i < i'$ and thus $|U|-1-i > |U|-1-i'$, and the dimension $\cdimnv{j}$ sum falls below the minimum.
\end{itemize}
That is, the addition of $\keyfunc{S_1}$ creates an ``$a_i$-shaped $j$-colored surplus'' that can be utilized by removing a set containing $a_i$ (but no other $j$-colored element).
Thus a set tuple $\tupfunc{S_2}$ may only be removed if $S_2 \subseteq S_1$.
Futhermore, since $C_1$ was an exact cover, only one set covers any given element.

After removing set tuples, set tuples may also be added.
The maximum value of $\cdimc{j}$ for each $j \in [23]$ (and aforementioned prevention of removing any elements not in $S_1$) prevents covering any element in $U$ by more than one set (tuple).

Since the sum value in each color value dimension was the minimum value when $\keyfunc{S_1}$ was added, removing $\keyfunc{S_1}$ is only possible if the configuration has these sums again.
This only occurs if the reconfiguration has restored the exact covering.
Any exact cover obtained by replacing a partition of $S_1$ with another partition of $S_1$ is reachable from $C_1$ via a union (into $S_1$) followed by a partition. 
\end{proof}

%Note that for proving Theorem~\ref{thm:paths} with $d=O(n)$ instead, we can simply consider the 0-1 LP formulation of the independent set problem on planar graphs, whose reconfiguration version is known to be $\PSPACE$-complete~\cite{KMM12}.
%\footnotesize

\paragraph{Acknowledgements.}
This work was initiated at the 32nd Bellairs Winter Workshop on Computational Geometry, January 27-February 3, 2017. We thank the other participants of the workshop for a productive and positive atmosphere. 

\bibliographystyle{plain}
\bibliography{paper}

\end{document}